\documentclass[structabstract]{aa}  
\usepackage{graphicx}
\begin{document}

   \title{{\bf The jet of the BL Lacertae object PKS 2201+044: MAD near-IR adaptive optics observations and comparison with optical, radio and X-ray data.}}
   \subtitle{}

\author{E. Liuzzo\inst{1,2}, R.Falomo\inst{3}, A. Treves\inst{4}, D. Donato\inst{5},  M. Sambruna\inst{5}, C. Arcidiacono\inst{3}, G. Giovannini\inst{1, 2}, J. Farinato\inst{3},A. Moretti\inst{3}, R. Ragazzoni\inst{3}, E. Diolaiti\inst{6}, M. Lombini\inst{6},  R. Brast\inst{1}, R. Donaldson\inst{7}, J. Kolb\inst{7}, E. Marchetti\inst{7} and S. Tordo\inst{7} }

\institute{Istituto di Radioastronomia, INAF, via Gobetti 101, 40129 
Bologna, Italy.\\
\email{liuzzo@ira.inaf.it}
\and
Dipartimento di Astronomia, Universit\`a di Bologna, via Ranzani 1 , 40127 Bologna, Italy.
\and
Osservatorio Astronomico di Padova, INAF, vicolo dell'Osservatorio 5, 35122 Padova, Italy.
\and
Universit\`a dell'Insubria (Como), Italy, associated to INAF and INFN.
\and
NASA Goddard Space Flight Center, Code 661, Greenbelt, MD 20771.
\and
Osservatorio Astronomico di Bologna, INAF, Bologna, Via Ranzani 1, 40127 Bologna, Italy.
\and
European Southern Observatory, Karl-Schwarschild-Str 2, 85748 Garching bei M\''{u}nchen, Germany
}

\date{Received ... accepted ...}

  \abstract
   {  Relativistic jets are a 
common feature of radio loud active galactic nuclei. Multifrequency observations are a unique tool to constrain their physics.} 
   { We report on a detailed study of the properties of the jet of the  nearby BL Lac object PKS 2201+044, one of the rare cases where the jet is detected from radio to X-rays.} 
   {We use new adaptive optics near-IR observations of the source,
obtained with the ESO multi-conjugated adaptive optics demonstrator (MAD) at the Very Large Telescope. These observations  acquired in Ground-Layer Adaptive Optics mode are combined with images previously achieved by HST, VLA and Chandra to perform a morphological and photometric study of the jet.}
   { We find a noticeable similarity in the morphology of the jet at radio, near-IR and optical
   wavelengths. We construct the spectral shape of the main knot of jet that appears dominated by synchrotron radiation.}
   { On the basis of the jet morphology and the weak lines spectrum we suggest that PKS 2201+044 belongs to the class of radio sources intermediate between FRIs and FRIIs.}

   \keywords{galaxies: BL Lacertae objects: individual: PKS 2201+044 - instrumentation: adaptive optics               }

   \maketitle

\section{Introduction.}

Radio loud active galactic nuclei (AGN) are characterized by the presence of relativistic jets mainly detected in radio band. In some cases, the jets can also be observed in the visible and at higher frequencies. In particular, this is so for some nearby BL Lac objects. They represent a subclass of AGN characterized by weakness of lines, luminous, rapidly variable  non-thermal continuum, significant polarization, strong compact flat spectrum radio emission, and superluminal motion (e.g. Ulrich et al. 1997, Giroletti et al. 2004).
Similar properties are also observed in flat
spectrum radio quasars and these two types of AGN are often grouped together
into the class of Blazars. Additional evidence of the  presence of relativistic jets in BL Lacs is given by their strong gamma emission (Abdo et al. 2010).

However, because of the close alignment of the jet with the line of sight, it is very difficult
 to distinguish it unless the angular resolution is sufficiently high. In fact, the detection of the jet at the
 different frequencies depends both on the limited angular resolution and
 on the short lifetime of the high energy electrons producing the non- thermal
 emission of the jet. 

The advent of high resolution imaging facilities such as HST and Chandra allowed
systematic searches for angular resolved multiwavelength counterparts of radio
jets in Blazars (Scarpa et al. 1999a) and also the detection of jets in radio-loud quasars at high redshift (Schwartz et al. 2000, Sambruna et al. 2002, Tavecchio et al 2007).  Moreover, the use of adaptive optics assisted imaging on Large Telescope started to permit to investigate from the ground the near-IR jet properties (Hutchings et al. 2004, Falomo et al. 2009).

In this paper we present near-IR images acquired with an innovative 
adaptive optics (AO) camera of PKS 2201+044, the nearest BL Lac object (z= 0.027). Morphological and photometric properties are discussed together with the observations in radio, optical, and X-ray bands (Scarpa et al. 1999b, Sambruna et al. 2007).

We adopt the concordance
cosmology with $\Omega_m$=0.3, $\Omega_\lambda$=0.7 and H$_{0}$= 70 km s
$^{-1}$ Mpc$^{-1}$. In this scenario, at z=0.027 1 arcsec corresponds to 0.54 kpc.
We defined $\alpha$ as S $\sim$$\nu^{-\alpha}$ where S is the flux density at frequency $\nu$.

\section{PKS 2201+044} \label{pks}

This source was originally classified as a BL Lacertae object (Weiler $\&$ Johnston 1980). Later spectroscopic studies found the characteristic absorption features of an elliptical galaxy. Its spectrum shows also the presence of weak broad and narrow H$_{\beta}$ (V\'eron-Cetty $\&$ V\'eron 1993) and H$_{\alpha}$ (Falomo et al. 1987, Sbarufatti et al. 2006) emissions together with [NII]$_{\lambda 6548, 6583}$, [S II] $_{\lambda 6717, 6713}$ (Sbarufatti et al. 2006). 
The morphology of the host galaxy is consistent with a luminous (M$_{V}$ $\sim$ -20.8 ) giant elliptical (Falomo et al. 1993).

Extensive radio observations of jet and halo were collected by Laurent-Muehleisen et al. 1993.
The radio source has a total extent of 5.6 arcmin
(Ulvestad $\&$ Johnston et al. 1984) and it has a compact ($\leq$ 3 mas) core as revealed by VLBI observations at 2.29 GHz (Preston et al. 1985). The jet emerges from the core as a narrow, collimated stream where two bright enhancement regions (knot $a$ and $b$ in Fig. 3) and one final radio trident-like structure (knot A) are visible (Sambruna et al. 2007). 
At parsec scale, in VLBI
images (Kollgaard et al. 1995) the source shows a bright core with a single jet
component. 

In X- ray images (Sambruna et al. 2007), an
emission structure is found close to the nucleus that is
similar to the other FRI sources detected with \textit{Chandra}. Overall the X-ray jet exhibits intermediate properties between low- and high-power radio sources.
The jet and the circumnuclear properties are similar to
those of the other FRI sources studied with Chandra, while the nuclear properties are
unusual showing narrow and broad optical lines. It is possible that for this
source we are seeing its jets at closer angles than in classical FRIs (Sambruna et al. 2007).

In the optical band, the emissions from knot $a$ and $b$ (Fig. 2) are weak while feature A is bright (Scarpa et al. 1999, Sambruna et al. 2007). Counterparts of knot A and knot $a$ are faintly revealed also in Chandra X-ray images (Sambruna et al. 2007).

\section{Observations and data analysis.}

\subsection{MAD near-IR data.}

We performed Ks-band observations of PKS 2201+044 on September 26 2007 using the European Southern
Observatory (ESO) Multi-Conjugate Adaptive Optics Demonstrator (MAD), mounted at UT4 (Melipal) of the Very
Large Telescope (VLT).
MAD is a prototype Multi Conjugate Adaptive optics (MCAO) system which aims to demonstrate
the feasibility of different MCAO reconstruction techniques in
the framework of the E-ELT concept and the 2nd Generation VLT Instruments. MAD is designed
to perform wide Field of View (FoV) adaptive optics correction in J, H and Ks band over
2 arcmin on the sky by using relatively bright (m$_{v}$ $<$ 14) Natural Guide Stars (NGS).

We refer to Marchetti et al. 2003 for the detailed description of MAD.
The MCAO correction was obtained using Layer Oriented Multi-Pyramid Wave Front Sensor for the Layer Oriented MCAO reconstruction (Ragazzoni 1996, Ragazzoni et al. 2000, see also Marchetti et al. 2005).
The detector has a $57\hbox{$^{\prime\prime}$ }\times57\hbox{$^{ \prime\prime}$ }$ field of view with pixel size of 0 $\!\!^{\prime\prime}$.028.

During the observations, we used three reference stars (Fig. \ref{field}). The seeing measured by the ESO DIMM in the V band (rescaled at the Zenith) during the
observation was $\sim$ 1 arcsec, with an rms of $\sim$0.05 arcsec. The elevation was within $\sim$ 55.5 and 60.7 degrees, corresponding to 1.21 and 1.15 airmasses. We obtained a total
of 5 dithered groups of images, for an overall number of 45 exposures along a pattern that covered a region of $\sim$ 77.3$\times$ 77.3 arcsec$^{2}$. 
Because of a problem with the internal optical derotator during this observation the close loop was not stable in MCAO mode. These data were therefore obtained with the partial adaptive optic correction (Ground-Layer Adaptive Optics mode). In spite of this inconvenience we are able to secure images with quality of FWHM smaller than 0.3 arcsec under non optimal seeing condition (see Table 1).
The total integration time was 2700~s. A summary of the observations and data analysis is presented in Table~\ref{tabblock}.

\begin{table}
\caption{ Journal of MAD observations.}\label{tabblock}
\begin{center}
\begin{tabular}{c c c c c }
\multicolumn{5}{l}{ } \\ \hline \\
Dither position & UT Time & Seeing$^{(1)}$ & T$_{exp}$ & FWHM$^{(2)}$ \\
                &         & arcsec &  sec      & arcsec \\
\hline
\hline
\\
a) & 01:26:40.3183 & 1.15 & 540 & 0.31 \\
b) & 01:42:03.7260 & 0.98 & 540 & 0.25 \\
c) & 01:56:06.8982 & 1.03 & 540 & 0.23 \\ 
d) & 02:10:12.1379 & 0.98 & 540 & 0.23 \\
e) & 02:22:53.3385 & 0.93 & 540 & 0.32 \\
All & --  & 1.01 & 2700 & 0.26 \\
\hline \\
\multicolumn{5}{l}{ $^{(1)}$The seeing measured by the ESO-DIMM in the V band and } \\
\multicolumn{5}{l}{ rescaled at the Zenith} \\
\multicolumn{5}{l}{ $^{(2)}$ For the FWHM refer to Sec. 3.1}
\end{tabular}
\end{center}
\end{table}

\subsection{Data reduction and calibration.}

Data reduction for these new near-IR observations was performed following the same procedure used for PKS 0521-365 in Falomo et al. 1999. In brief, we applied standard data reduction procedures for near-IR observations, which include
trimming of the frames, bad pixel masking, dark and flat-field correction, and
sky-background subtraction. The individual sky subtracted and re-aligned 
images were stacked into the final combined image of the field. For the flat-field correction, we used the median of several images
obtained on the sky at the beginning of each night.  
Then we constructed the reference sky image from the median of all science
images of a given observation. The region around the target was excluded from the evaluation of the sky 
during the process of combining dithered images.
We further checked that the residual of the faint extended emission 
from the host galaxy is not detected in the final sky image 
(that is order of magnitude brighter than the faint light from the galaxy). In the case of this target, the field is not
affected by crowding of stellar sources, which allows for the creation of a good
quality sky image.  This was then normalized to the median counts and subtracted.
Finally, we combined all sky-subtracted science images after properly aligning
them using as a reference the positions of the brightest (non saturated) sources
detected.
From the final co-added images we measured the associated image quality from the
average full width at half maximum (FWHM) of the intensity profile of a number of
reference stars.
Photometric standard stars (Landolt 1993) were observed during the night and used 
 to calibrate MAD frames, and an accuracy of 0.1 magnitude is
derived. Astrometric calibration was obtained using stars in the field allowing an accuracy of $\sim$0.3 arcsec.

\begin{figure}[ht!]
\centering
\includegraphics[width=8cm]{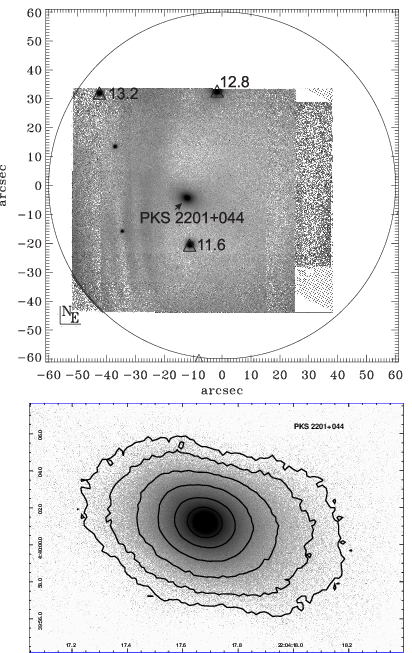}
\caption{ Top: The 2 arcmin Field of View (black circle) corrected by the MAD
MCAO system. The picture is the mosaic of all the images collected with the
camera. The three triangles correspond to the Layer Oriented AO reference stars.
R-magnitude are reported for each star. The imaged area 
is 77.3$\times$89.9 arcsec$^{2}$. The imaged area is 77.3$\times$89.9 arcsec$^{2}$. 
Bottom: Enlargement of central part zooming on PKS 2201+044.} \label{field}
\end{figure}

The artifact visible in left side of the detector (see Fig.\ref{field}) is due to imperfect baffling of the stray light of the instrument at the time of the observations. Since we  centered the target in the right part of the detector these artifacts have negligible effects of our final image of the jet and its surrounding area.     

 An enlargement of the central part of the MAD is given in Fig. \ref{field}, bottom panel.

\begin{figure} [ht!]
\centering
\includegraphics[width=5cm]{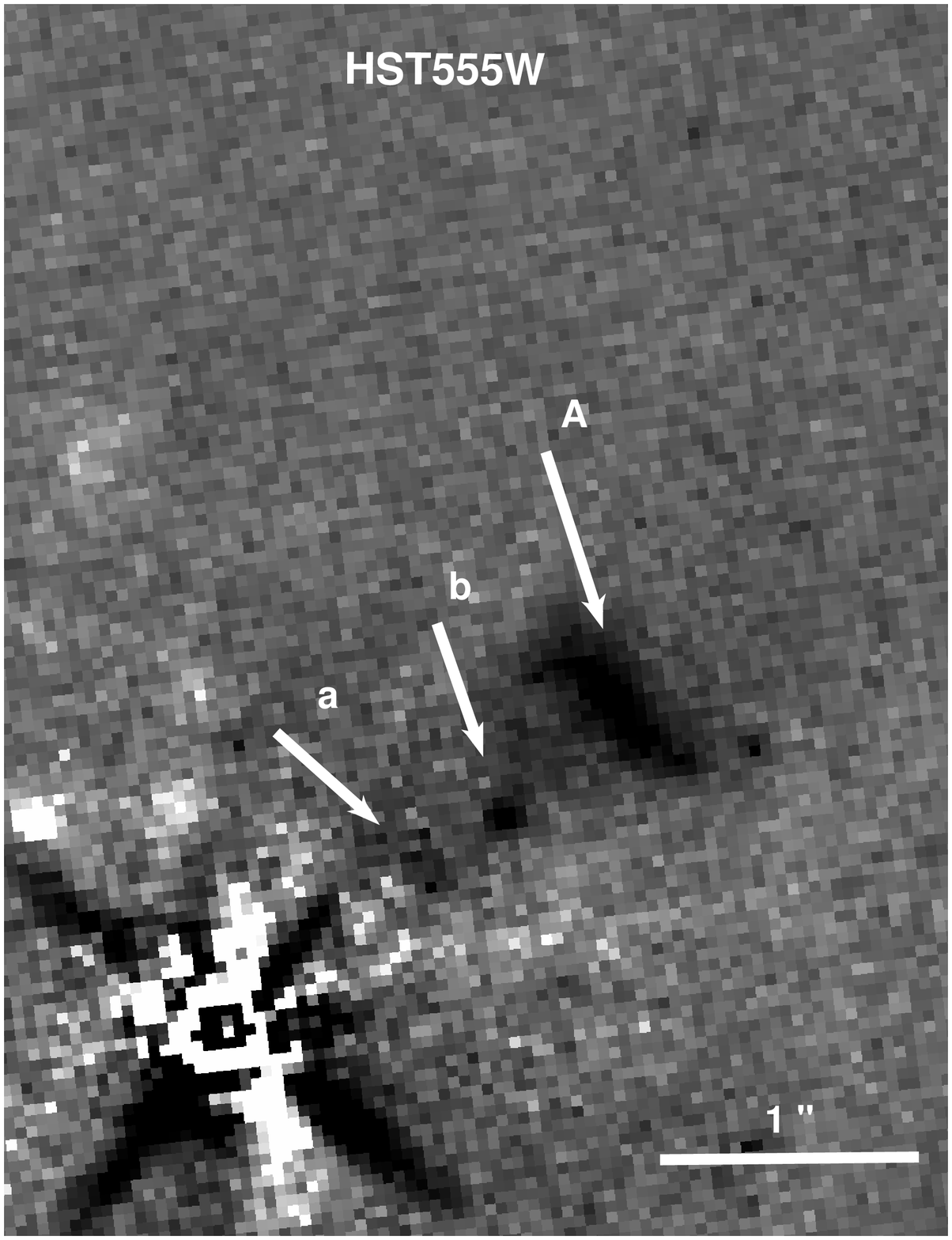} 
\hfill
\includegraphics[width=5cm]{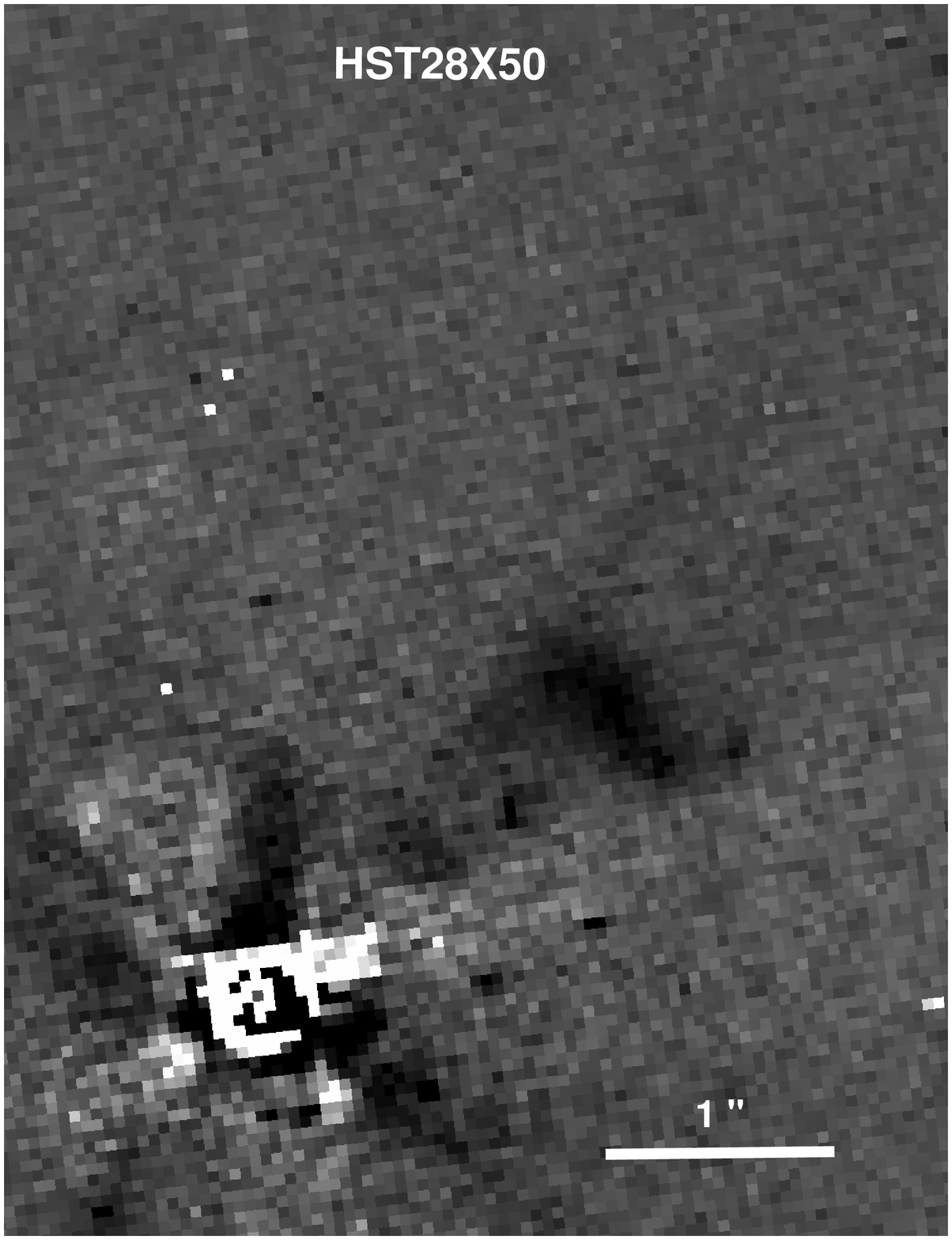} 
\hfill
\includegraphics[width=5cm]{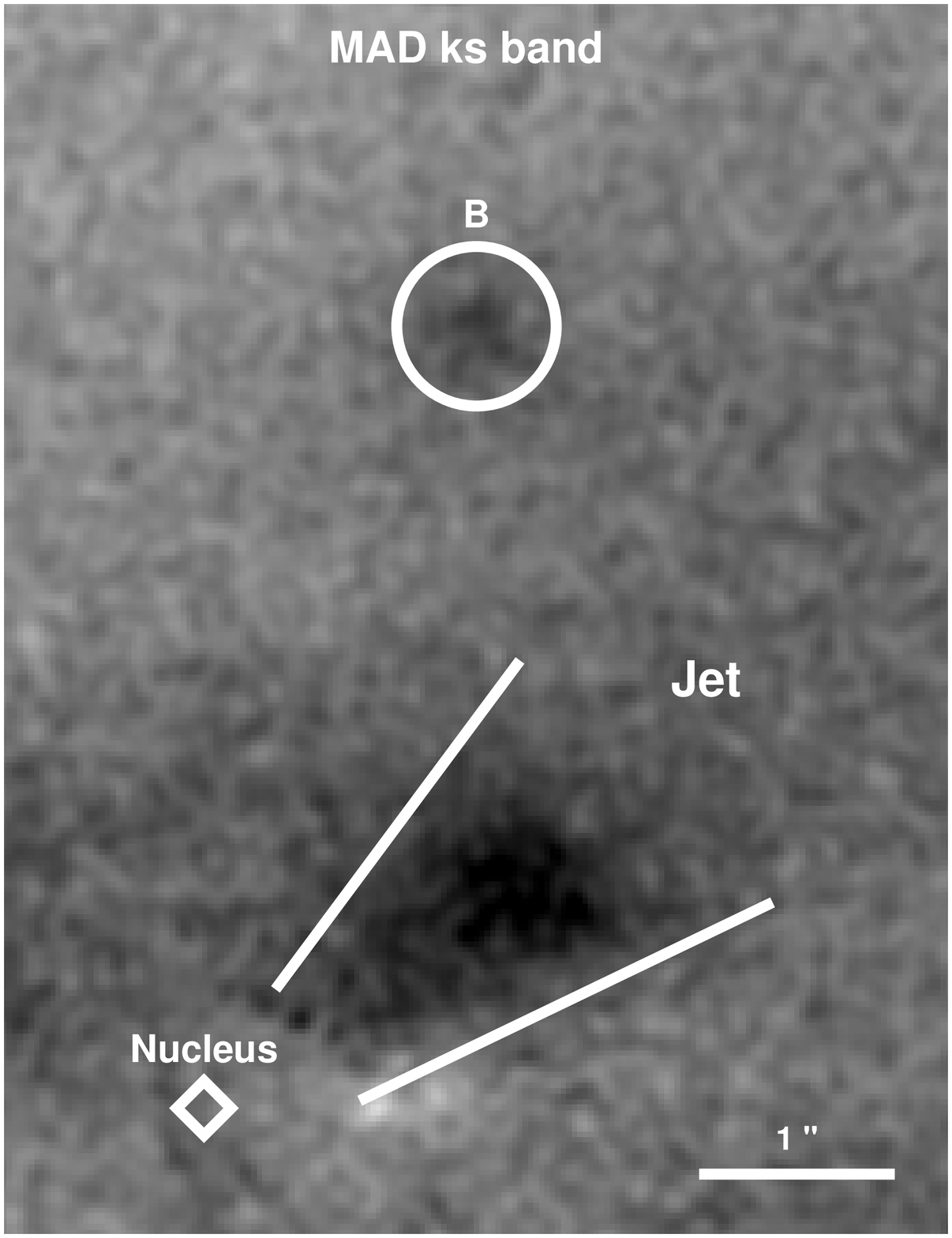}
\caption{ Top panel: HST F555W image from data of Sambruna et al. 2007. Knot $a$, $b$ and A are indicated. Middle panel: HST 28X50LP image from Sambruna et al. 2007. Bottom panel: near-IR map of the jet of our object as observed by MAD at VLT in Ks filter (see Sec. \ref{sect_rio}). The nucleus, the elongation of jet and target B are marked. The orientation of the jet is SE-NW respect to the nucleus. Shown field of view is 4.7 $\times$ 6.2 arcsec $^{2}$. The optical and near-IR images have a spatial resolution of $\sim$0.15 \arcsec and $\sim$0.2 \arcsec respectively. } \label{fig.jet}
\end{figure}

 In order to study the jet itself, we proceed as in Falomo et al. (2000), building a model of the
host galaxy and its nucleus and subtracting it to the original MAD images. For the inner
part, the model of the PSF was derived from the field stars. After masking the regions containing the jet, the isophotes of the host galaxy were fitted by
ellipses. The remaining signal (the jet) is shown in Fig. \ref{fig.jet} where we also report the HST F555W and F28X50LP ($\lambda$ = 7218 \AA) images taken from Sambruna et al. 2007. The faint extended nearly horizontal emission outside the jet region is an artifact generated during image sky subtraction. We took into account it and subtracted in our flux estimations (Sect. \ref{sec.res}). Finally, in our MAD observations, at the north of the jet, at $\sim$ 4.3 arcsec from the nucleus, there is a detection of another source (target B in Fig. \ref{fig.jet}) with a total magnitude in Ks band of 21.7. This object does not appear in HST images (Fig. \ref{fig.jet}).  We searched in optical, radio , infrared and X-ray source catalogs but no identification of this feature was found. It is likely a faint background red galaxy.

\section{Results.} \label{sec.res}

\subsection{ Jet morphology: radio, NIR and optical comparison.} \label{sect_rio}

In Fig. \ref{fig.jet} it is reported the MAD image for the jet emission. The jet shows a diffuse conical morphology with a total flux of 25 $\mu$ Jy.
Comparing the near-IR with the optical jet emission (Fig. \ref{fig.jet}), some difference is apparent. Unlike the optical maps, in the near-IR image only the knot A is present. Moreover, in the optical wavelengths, as revealed by HST F555W and HST 28X50LP images, knot $b$ seems to be bluer being more brilliant in HST F555W. 

\begin{figure} [ht!]
\centering
\includegraphics[width=8cm]{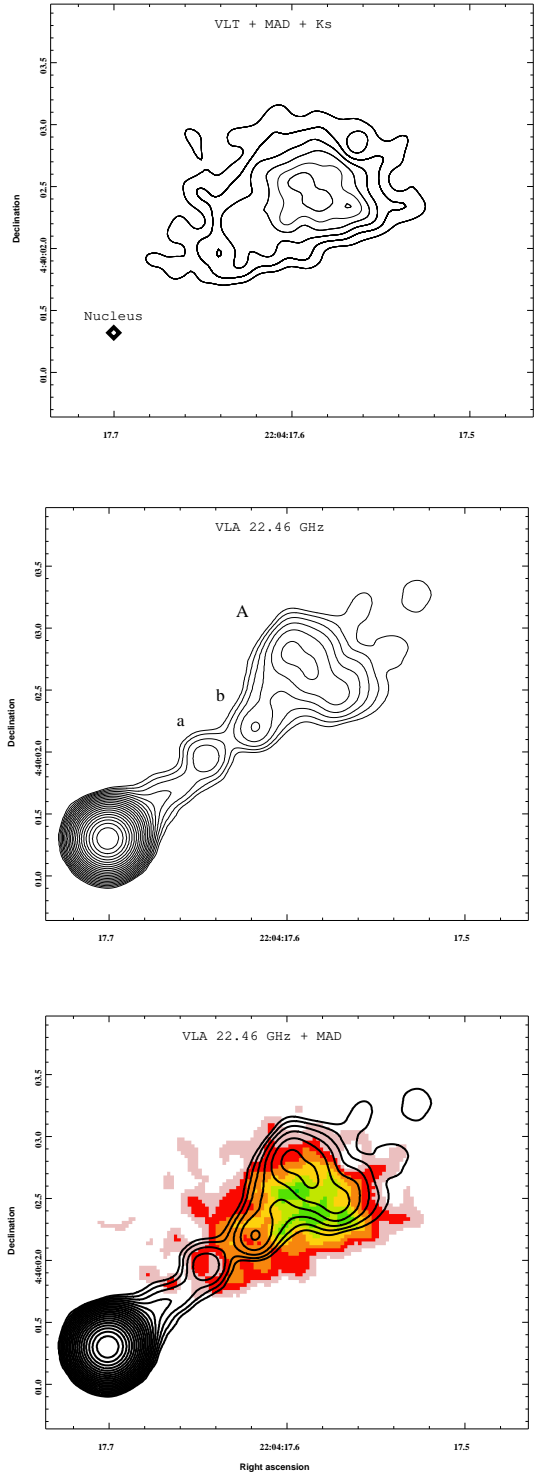} 
\caption{Top panel: Contour plot of the jet emission of PKS 2201+044 observed by MAD in the Ks band. The diamond represents
the position of the core. Middle panel: Image of the jet observed in radio band at 22.46 GHz with VLA (Sambruna e tal. 2007). The
resolution is 0.25 arcsec. Contour levels are 0.2, 0.3, 0.4, 0.6, 0.8, 1, 2, 4, 8, 12, 15, 30, 50, 60, 100 and 200 mJy/beam.
The peak level is 0.28 Jy/beam. Bottom panel: NIR jet emission (colour scale) of PKS 2201+044 overimposed to VLA radio
emission at 22.46 GHz (contour map).}\label{rir}
 \end{figure}

In Fig.\ref{rir} we compare the MAD near-IR image  of PKS 2201+044 with the VLA radio map at 22.46 GHz previously obtained by us (Sambruna et al. 2007). The overall morphology of the jet in the near-IR is similar to that observed in the radio band.

In order to give an appropriate comparison of the jet morphology in radio and NIR wavelengths, we construct longitudinal and transverse brightness jet profiles (see Figs. \ref{jetprofl}-\ref{jetproft}). We used the new near-IR data and maps at 22.46 GHz obtained with VLA (Sambruna et al. 2007) that have comparable spatial resolution.
 
Jet profiles are extracted integrating the signal in a box of $\sim$ 1.3 arcsec width  and 1 arcsec width for the longitudinal and transverse profile respectively. The box sizes are the best compromise between an increase of the signal to noise without loss of information on the possible jet substructures due to the integration. 

The longitudinal profiles, at both frequencies, show a peak of emission around 2.2 arcsec from the core that corresponds to knot A. Knots $a$ and $b$, respectively at $\sim$ 1 arcsec and $\sim$ 1.5 arcsec from the core, are resolved in the radio profile (Fig.\ref{jetprofl}), while in the near-IR longitudinal profile no clear secondary peaks are visible.
The width of the knot A profile in the radio map is smaller than this in MAD map ($\sim$ 0.6 arcsec versus $\sim$1 arcsec). 

In Fig.\ref{jetproft}, we show the result of a transverse profile for the knot A. No particular structures are observed. Similarly, there is coincidence between the emission peak at both frequencies. 
Unlike the longitudinal profile at the beginning of the jet (see Fig. \ref{rir}), knot A transverse profile has the similar width ($\sim$1.2 arcsec) in radio and near-IR data (Fig.\ref{jetproft}).

We look also at the radio morphology at different frequencies and resolutions (Fig. \ref{2cc}).
At low resolution (i.e VLA 8.4 GHz), within 3.5 arcsec from the nucleus, the target appears double with one component that corresponds to the core (C) and the second (A) one that shows the peak of the emission coincident with the peak of the knot A observed at higher resolution. As discussed before, intermediate resolution observations (i.e. VLA 22.46 GHz) allow us to resolve the nucleus and three components.
High resolution data were obtained by us (Sambruna et al. 2007) with MERLIN observations at 5 GHz (Fig. \ref{peak}). At this resolution ($\sim$0.1 arcsec), the whole source shows a total flux of $\sim$ 350 mJy and a total elongation of $\sim$ 3.5 arcsec in direction SE-NW.  
In the knot A, we note a peak of emission, A1 (Fig.\ref{peak}). A hint of this feature can be present also in the VLA data (Fig. \ref{jetproft}). Studies of Sambruna et al. 2007 show the presence of a complex polarization structure on the flaring point knot A. We suggest that this behavior could be explained as the result of the jet interaction with the surrounding medium. The structure begins collimated and in knot A the jet interacts with the external material making the radio emission more diffuse with a consequent deviation from the initial direction.

\subsection{ Spectral energy analysis.}\label{sedsec}

\begin{figure*} 
\centering
\includegraphics[width=10cm]{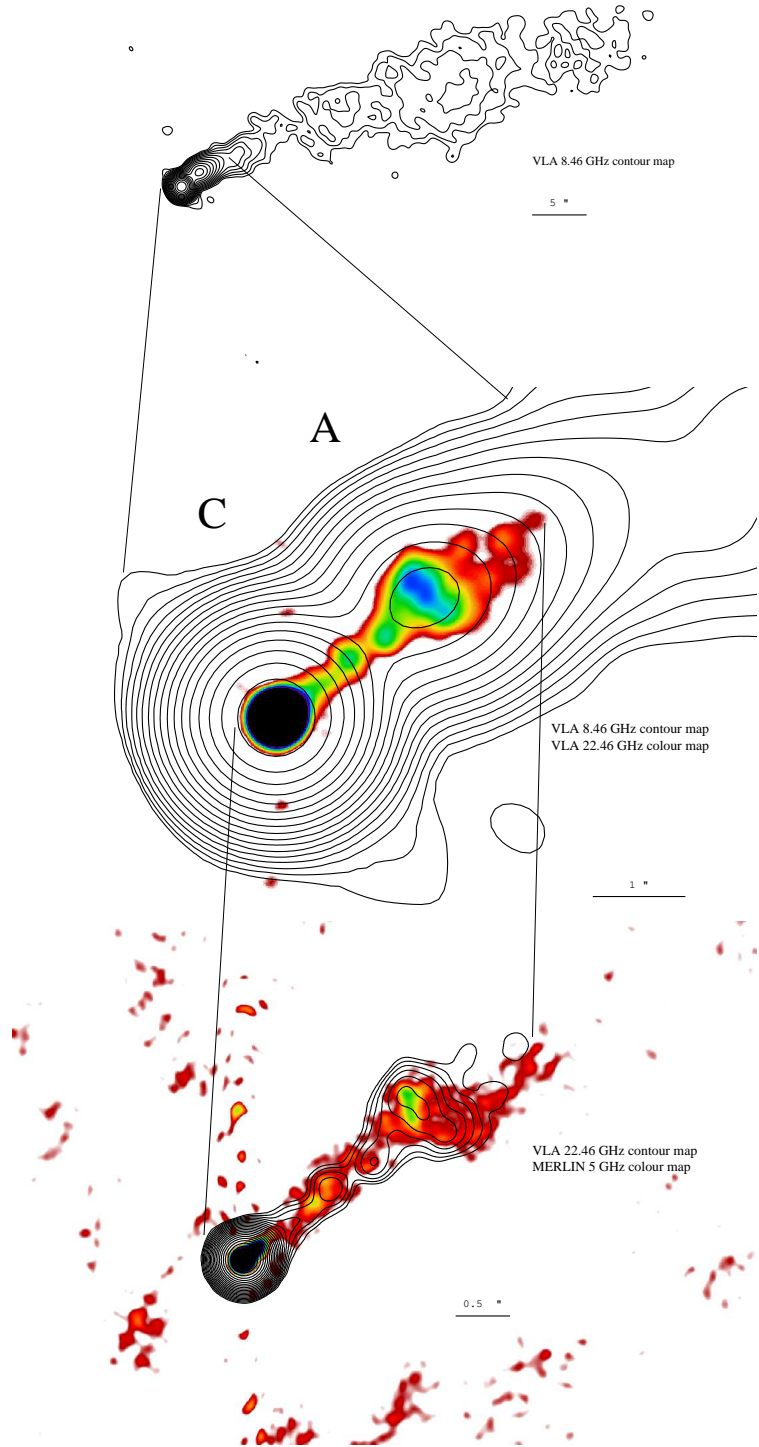}
\caption{Radio maps of the BL Lac object PKS 2201+044 at different frequencies and resolutions obtained by MERLIN and VLA. In each panel, labels report the type of data that are plotted and scale bars give an indication of the spatial resolution of the map.} \label{2cc}
 \end{figure*}

  We modelled the spectral energy distribution (SED) for A and C (Fig. \ref{2cc}) taking flux measurements reconsidering VLA data of Sambruna et al. 2007 at 1.41 GHz , 5 GHz, 8.4 GHz (plotted in Fig.\ref{2cc}) and 22.46 GHz at the same spatial resolution of 1 arcsec.  We used Synage Software that allows us to test different models of synchrotron emission. We considered different models: KP (Kardashev-Pacholczyk), JP (Jaffe-Perola), CI (Continuous Injection), CIE (Continuous Injection+adiabatic expansion), CIER (Continuous Injection+reduced adiabatic expansion) (Murgia et al. 1999 and references therein). The one that gives the best fit (based on $\chi^{2}$ evaluation) is the CI model. This assumes that the radio source evolution is described by a continuous injection model, where the sources are continuously replenished by a constant flow of fresh relativistic particles with a power law energy distribution, with exponent $\delta$. Under these assumptions, the spectrum has a standard shape (Kardashev et al. 1962), with a spectral index of injection $\alpha_{inj}$= ($\delta$-1)/2 below a critical frequency $\nu_{break}$ and $\alpha_{h}$=$\alpha_{inj}$ + 0.5 above $\nu_{break}$.  We found an $\alpha_{C,inj}\sim$0.38  and a break frequency $\nu_{C,break}\sim$1.3$\times$10$^{12}$ Hz for the component C and $\alpha_{A,inj}\sim$0.38  and a break frequency $\nu_{A,break}$$\sim$10$^{12}$ Hz for the component A. Using the method of the break frequency (Murgia et al. 1999), if we have the value of magnetic field B, Synage gives an estimate of the synchrotron age for C and A. Using the formula of Govoni et al. 2004, we determined for A and C the value of B in equipartition model and synchrotron emission hypothesis:
 \begin{center}
B = $(\frac{24\pi}{7} u_{min})^{0.5}$ where 
$u_{min}(erg/cm^{3})$ = $\xi(\alpha, \nu_{1}, \nu_{2}) (1 + k)^{4/7}$$ (\nu_{0}(MHz))^{4\alpha /7}$ (1 + z)$^{(12 + 4 \alpha)/7}$ I$_{0}$(mJy/arcsec$^{2}$)$^{4/7}$(d(kpc))$^{-4/7}$ 
\end{center}
 where z is the redshift of the source , I$_{0}$ is the source brightness at the frequency $\nu_{0}$, d is the source depth and the constant $\xi(\alpha, \nu_{1},\nu_{2})$is tabulated in Table 1 of Govoni et al. 2004. 
Similarly to the results for 3C 371 (Sambruna et al. 2007), we found B$_{C}$$\sim$80$\mu$G for C and B$_{A}\sim$69$\mu$G for A. From these values, Synage derives an estimate for the synchrotron ages of  6.1$\times$10$^{-2}$Myrs for C and  8.6$\times$10$^{-2}$Myrs for A consistently with the fact that the core contains the new ejected emission. 

Then, we combined our near-IR measurements of knot A with available optical,
radio and X-ray fluxes from Sambruna et al. 2007 to construct the SED. In agreement with Sambruna et al. 2007, we extracted NIR MAD flux measurement in a box with size of 0.3 arcsec$^{2}$. 
Fig.\ref{sed} shows the SED for the knot A from radio to X-rays. Its shape appears similar to that of a knot of 3C371 (Sambruna et al. 2007) for which it is suggested that the emission derives from a single synchrotron component following a power law from the radio to the optical band, peaking close to the optical wavelength range and softening between the optical and X-rays. 
We assumed that also for PKS 2201+044 the SED is described by synchrotron emission from the radio to the X-ray band and we fitted it using Synage Software.  For knot A, from all the available data from X to radio bands, we found $\alpha_{inj, tot}\sim$0.74 and $\nu_{break, tot}\sim$3.15$\times10^{12}$Hz. 

Finally, we made spectral index analysis of A1 using available radio multifrequency observations (Sambruna et al. 2007).  We used MERLIN map at 1.41 GHz, MERLIN data at 5 GHz and VLA map at 22.46 GHz. 
It results globally a steep spectral index: in particular, we found between 1.41 GHz and 5 GHz $\alpha_{1.41-5}\sim$0.82$\pm$0.04 and $\alpha_{5-22.46}\sim$0.5$\pm$0.03 between 5 GHz and 22.46 GHz. Steep spectral indexes are typical of emission for losses due to the interaction of the jet with the surrounding medium (Laing $\&$ Bridle 2002) confirming our hypothesis on A1 proposed in sect. \ref{sect_rio}.

\section{Conclusions.}

\begin{figure}
\centering
\includegraphics[width=9cm]{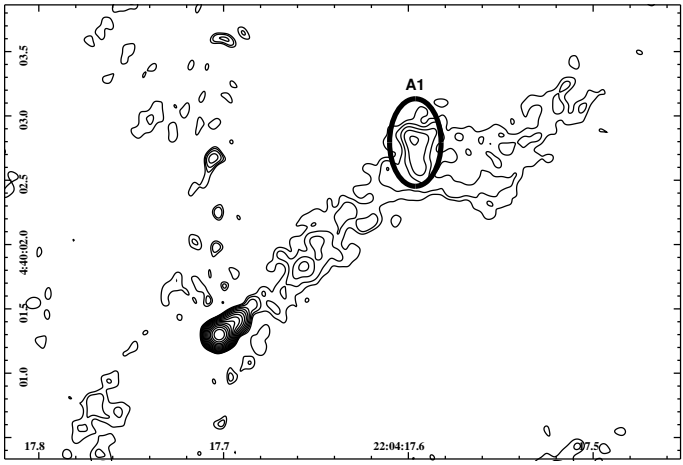}
\caption{ Contour plot of the jet radio emission of PKS 2201+044 observed with MERLIN at 5 GHz (Sambruna et al. 2007). The resolution is 0.1 arcsec. Contour levels are logarithmic beginning from  0.4 mJy/beam. The peak emission is 0.23 Jy/beam. For discussion on region A1 see Sec.4.1.} \label{peak}
 \end{figure}

\begin{figure} [ht!]
\centering
\includegraphics[width=9cm]{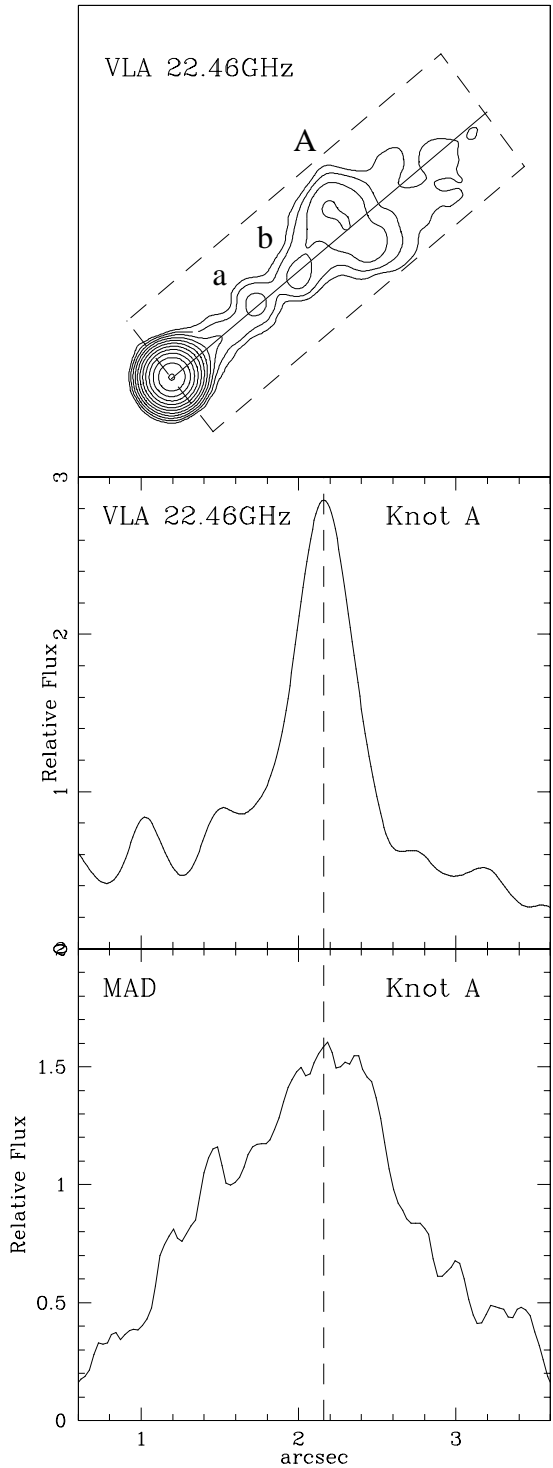}
\caption{ Longitudinal profile of the jet of PKS 22101+044 as observed in radio band by VLA at 22.46 GHz (middle panel) and by MAD at VLT in Ks band (bottom panel). The top panel shows the box (dashed line) in which the signal is integrated. The solid line represents the jet direction. The units of the y-axis
are counts pixels $^{-1}$ adequately scaled permitting an acceptable comparison between the frequencies analysed. The zero point corresponds to the core position. Knots $a$, $b$ and A are labeled. }\label{jetprofl}
 \end{figure}

\begin{figure}[ht!]
\centering
\includegraphics[width=9cm]{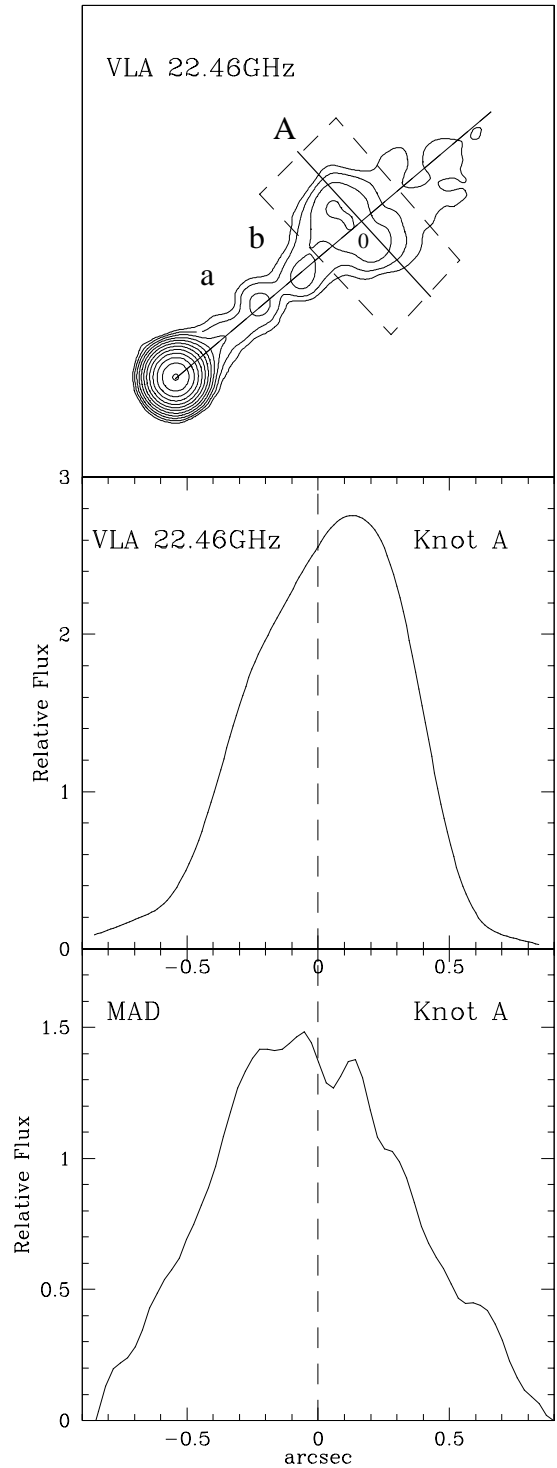}
\caption{Transverse profile of the jet of PKS 2201+044 as observed in radio band by VLA at 22.46 GHz (middle panel) and by MAD at VLT in Ks band (bottom panel). The top panel shows the box in which the signal is integrated. We consider as zero point the intersection between the jet direction and the middle line crossing the box (solid line). Negative values are in the SW direction. Knots $a$, $b$ and A are labeled. } \label{jetproft}
 \end{figure}

\begin{figure} [ht!]
\centering
\includegraphics[width=9cm]{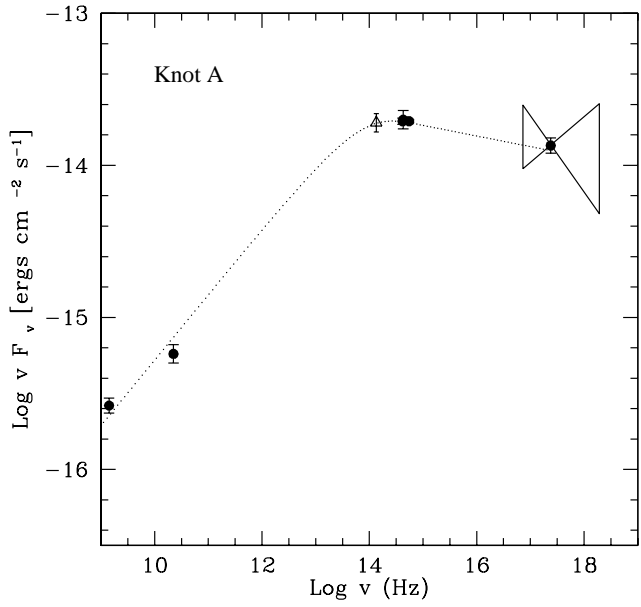}
\caption{SED of the knot A for PKS 2201+044. 
The open triangle represents the near-IR measurement from the MAD 
observation.  Radio, optical and X-ray data are taken
from Sambruna et al. 2007. The dashed line represents the best fit (see Sec. 4.3).} \label{sed}
\end{figure}

In this paper, we presented high resolution near-IR images of the jet of the nearly BL Lac object PKS 2201+044 using an innovative adaptive-optics device (MAD) built as a demonstrator for multi-conjugated AO imaging obtained with MAD-VLT instrument. These new data, together with the analysis of data previously obtained in optical (by HST) , radio (by VLA and MERLIN) and X-ray band, have provided us with insight into the jet associated with this nearby extragalactic source. The main results from this study are:\\i) the morphology of the jet is very similar at radio, near-IR and optical frequencies with knot A being the only substructure detected at all wavelengths and resolutions;\\ ii) near-IR MAD observations obtained in a GLAO mode give a good mapping of the diffusive emission intra-knots enhancing a conical shape;\\ iii) the emission from the knot A is dominated by single synchrotron component;\\ iv) we derived for the component A and C an estimate for the synchrotron age, 8.6$\times$10$^{-2}$ Myrs and 6.1$\times$10$^{-2}$ Myrs respectively;\\ v) a detailed study (Fig. \ref{peak}) reveals that knot A emission is not homogeneous, in particular in direction SW-NE.

This multiwavelength study of the jet of PKS 2201+044 shows strong similarity with the jets of two other well studied BL Lacs, 3C 371 (Sambruna et al. 2007) and PKS 0521-365 (Falomo et al. 2009).
 These objects are very peculiar being the only three BL Lacs for which the jet is detected in optical bands. It is worth to note that for all these objects weak broad emission lines are present in their optical spectra (Sbarufatti et al. 2006, Falomo et al. 2009, Sambruna et al. 2007). These suggest that they are intermediate sources between FR I and FR II, having multiwavelength FRIs properties (Ledlow $\&$ Owen 1996) but showing broad optical emission lines that FRI sources are generally inefficient to produce (Baumm et al. 1995). We argue that the presence of broad optical emission lines could be due to intermediate properties of their broad line regions. 
Future investigations of other similar objects showing jet would clarify this point.

\begin{acknowledgements}
We wish to thank Teddy Cheung for his work on radio data reduction. This work was supported by contributions of European Union, Valle D'Aosta Region and the Italian Minister for Work and Welfare.
This research has made use of the NASA/IPAC Extragalactic Data Base (NED), which is operated by the JPL, California Institute of Technology, under contract with the National Aeronautics and Space Administration.
\end{acknowledgements}

\clearpage

\clearpage

\clearpage

\clearpage

\clearpage

\clearpage
cd ..

\end{document}